\documentclass[prb,aps,showpacs,twocolumn,amsmath,amssymb,superscriptaddress,affilletter]{revtex4-1}
\usepackage{color}
\usepackage{graphicx}
\usepackage{tabularx}
\usepackage[T1]{fontenc} %polish fonts
\usepackage[cp1250]{inputenc} %polish fonts

\newcommand{\be}{\begin{equation}}
\newcommand{\ee}{\end{equation}}

\begin{document}
\title{Stark Spectroscopy and Radiative Lifetimes \\ in Single Self-Assembled
CdTe Quantum Dots}% Force line breaks with \\

\author{Ł.~Kłopotowski} \affiliation{Institute of Physics, Polish Academy of Sciences,
02-668 Warsaw, Poland}

\author{V.~Voliotis} \affiliation{Université Pierre et Marie Curie,
CNRS, 4 Place Jussieu, 75252 Paris, France}

\author{A.~Kudelski} \affiliation{CNRS-Laboratoire de Photonique et
Nanostructures, Route de Nozay, 91460 Marcoussis, France}

\author{A.~I.~Tartakovskii} \affiliation{Department of Physics and Astronomy, University of Sheffield, S3 7RH, United Kingdom}

\author{P.~Wojnar} \affiliation{Institute of Physics, Polish Academy of Sciences,
02-668 Warsaw, Poland}

\author{K.~Fronc} \affiliation{Institute of Physics, Polish Academy of Sciences,
02-668 Warsaw, Poland}

\author{R.~Grousson} \affiliation{Université Pierre et Marie Curie,
CNRS, 4 Place Jussieu, 75252 Paris, France}

\author{O.~Krebs} \affiliation{CNRS-Laboratoire de Photonique et
Nanostructures, Route de Nozay, 91460 Marcoussis, France}

\author{M.~S.~Skolnick} \affiliation{Department of Physics and Astronomy, University of Sheffield, S3 7RH, United Kingdom}

\author{G.~Karczewski} \affiliation{Institute of Physics, Polish Academy of Sciences,
02-668 Warsaw, Poland}

\author{T.~Wojtowicz} \affiliation{Institute of Physics, Polish Academy of Sciences,
02-668 Warsaw, Poland}

\date{\today}% It is always \today, today,
             %  but any date may be explicitly specified
\begin{abstract}
We present studies on Coulomb interactions in single self-assembled CdTe quantum dots. We use a field effect structure to tune the charge state of the dot and investigate the impact of the charge state on carrier wave functions. The analysis of the quantum confined Stark shifts of four excitonic complexes allows us to conclude that the hole wave function is softer than electron wave function, i.\! e.\! it is subject to stronger modifications upon changing of the dot charge state. These conclusions are corroborated by time-resolved photoluminescence studies of recombination lifetimes of different excitonic complexes. We find that the lifetimes are notably shorter than expected for strong confinement and result from a relatively shallow potential in the valence band. This weak confinement facilitates strong hole wave function redistributions. We analyze spectroscopic shifts of the observed excitonic complexes and find the same sequence of transitions for all studied dots. We conclude that the universality of spectroscopic shifts is due to the role of Coulomb correlations stemming from strong configuration mixing in the valence band.
\end{abstract}

%\keywords{Suggested keywords}%Use showkeys class option if keyword
                              %display desired
\maketitle

%\textcolor[rgb]{1.00,0.00,0.00}{xxx I guess an Introduction is
%missing}

\section{Introduction}

In order to fully exploit the potential of quantum dots (QDs) in numerous proposed optoelectronic and quantum computing applications, a detailed knowledge of Coulomb interaction between carriers occupying a dot is essential. The description of these interactions is usually done in the language of Hartree-Fock (HF) approximation, in which the carrier wave functions are given by single orbitals, subsequently corrected by correlation terms originating from mixing of different orbital configurations (configuration interaction --- CI). Embedding QDs in vertical field effect structures provides the possibility of electrical control of the QD charge state\cite{dre94} and allows to study many fundamental properties related to Coulomb interactions. In particular, relative impact and magnitude of the direct HF and correlation terms on the spectroscopic shifts of QD transitions upon charging were evaluated\cite{reg01,bes03,dal08,cli08}. It was also shown that the dots can be fed with individual charge carriers one by one \cite{tar96} as each of them raises an electrostatic barrier, which needs to be overcome to add the next carrier. This Coulomb blockade results in a possibility of obtaining a charge-tunable device, in which the charging of a single dot is precisely controlled by external voltage and manifested by distinct charging steps in photoluminescence (PL) spectra \cite{war00,fin01,bai01,edi07}. In turn, the feasibility of preparing a dot in a given charge state leads to discoveries of effects particular to a given charge configuration such as creation of dynamical nuclear polarization \cite{ebl06,tar07}, two-qubit conditional quantum-logic operations \cite{ram08}, and electrical control of a spin state of a single Mn atom \cite{leg06}. Moreover, application of a voltage results in an electric field, which shifts QD transition energies due to a quantum confined Start effect (QCSE) \cite{emp97,ray98,seu01,fin04,kow06}. The shifts can be exploited as a resonant optical spectrum analyzer with a voltage-tunable sensitivity \cite{zre02}, which allowed to measure the lifetime-limited absorption transition linewidth \cite{sei05}. From the point of view of fundamental studies, QCSE is sensitive to the charge distribution in the dot and reflects dot morphology. Indeed, analysis of the Stark shifts allowed to determine the orientation of the electron-hole dipole in an InAs dot with GaAs barriers \cite{fry00} and interpret this effect as a result of an In/Ga intermixing during growth \cite{fin04,bar00}. Redistributions of carrier wave functions upon charging were investigated by time-resolved photoluminescence as the exciton lifetimes directly reflect the electron-hole overlap \cite{dal08,feu08}. It was found that upon charging holes undergo a stronger modification than electrons, owing to a weaker confinement in the valence band -- a result subsequently confirmed by a theoretical work \cite{cli08}. Moreover, it was shown that weaker confinement leads to larger oscillator strengths (i.e. shorter lifetimes) of excitons confined in monolayer fluctuation QDs.\cite{bel98,hou05}

Charge tunability is obtained routinely in III-V QDs and most of the reports cited above relate to the InGaAs system. On the other hand, the physics of Coulomb blockade in II-VI systems is relatively unknown. Charge tunability was demonstrated in CdSe QDs \cite{seu03} and then on a CdTe QD with a single Mn ion \cite{leg06}. QCSE was reported only for an electric field applied in the QD plane \cite{seu01}. These achievements notwithstanding, a lot of vital information on Coulomb interactions in II-VI systems is missing.

In this report, we present studies of Coulomb interactions in CdTe QDs by two methods. In Section III, we show PL studies of a QD embedded in a Schottky diode structure, where the charge state can be tuned from $-e$ to $+e$ by an application of a vertical bias. It allows us to access the charge distributions and their modifications under external electric field for various few-body complexes. Stark effect spectroscopy provides a direct access to electron-hole polarizability and permanent dipole moment --- both intrinsic parameters reflecting the form of electronic wave-functions. We address the redistribution of carrier wave functions upon charging and find that the electron is stiffer than the hole. We attribute this effect to stronger correlations in the valence band. This conclusion is further supported in Section IV, where we study PL lifetimes of different excitonic complexes. We find that the lifetime is more affected by an addition of an extra hole than by an extra electron. This correlation driven wave function redistribution results in a decrease of the electron-hole overlap and as a consequence, an increase in the transition lifetime. In Section V, we discuss the results in the language of configuration mixing and address the magnitudes of spectroscopic shifts recorded for the observed excitonic complexes. %We attempt a theoretical description of the exciton transition energies. We model the dot with a quantum disk with finite barriers and calculate the emission spectrum as a function of electric field. The calculations take into account the configuration mixing. We find that the correlation component dominates over the direct carrier - carrier interaction. However, we obtain a significant underestimation of the spectroscopic shifts upon charging and attribute it to uncertainty of the real dot morphology.

\section{Samples and Experiment}

The samples were grown by molecular beam epitaxy on a (100)-oriented GaAs substrate. Schottky diode structures contained a ZnTe buffer layer, $\sim 4 \mu$m thick, $p$-doped with nitrogen at a level of about 10$^{18}$ N acceptors per cm$^{3}$, which acted as a back contact. It was separated by an 80 nm wide intrinsic ZnTe spacer from a single layer of CdTe QDs. Dot formation was induced by changing the surface energy of a strained CdTe layer by deposition of amorphous tellurium \cite{tin03}. The above procedure yields approximately lens-shape dots, with base diameter in the range between 20 and 40 nm and heights between 2 and 8 nm. QD layer was capped by a 100 nm layer of ZnTe and another 100 nm layer of Zn$_{0.9}$Mg$_{0.1}$Te. The latter served as a blocking barrier to prevent the escape of carriers to the surface. On top, a semitransparent, 15 nm thick Ti/Au Schottky gate was deposited.

The cw PL signal was excited slightly below the ZnTe barrier with a 532 nm laser beam focused onto a $\sim 2 \mu$m spot. The estimated dot density is on the order of 10$^{10}$ cm$^{-2}$, so we excited roughly 500 dots. Individual dots were accessed by tuning the detection energy either to high- or low-energy tail of the inhomogenously broadened (FWHM $\sim$100 meV) PL band. The signal was detected by a nitrogen-cooled CCD camera coupled to a double monochromator.

Samples used for the time-resolved PL studies consisted of an undoped ZnTe buffer on top of which a single layer of CdTe QDs capped with another ZnTe barrier was deposited. To facilitate the access to single dots, 200 nm shadow mask apertures were processed by spin casting polybeads before metallization of a 100 nm gold layer.

For measurements of the PL lifetime, as an excitation source we used a frequency doubled output of an optical parametric oscillator pumped with a Ti:sapphire pulsed laser, yielding 2 ps pulses at 532 nm. The PL signal was time-resolved with a streak camera providing an overall temporal resolution of 10 ps. All measurements were performed at 10 K.

\section{Charge Tunability and Stark Effect}

\begin{figure}[!h]
  \includegraphics[angle=-90,width=.5\textwidth]{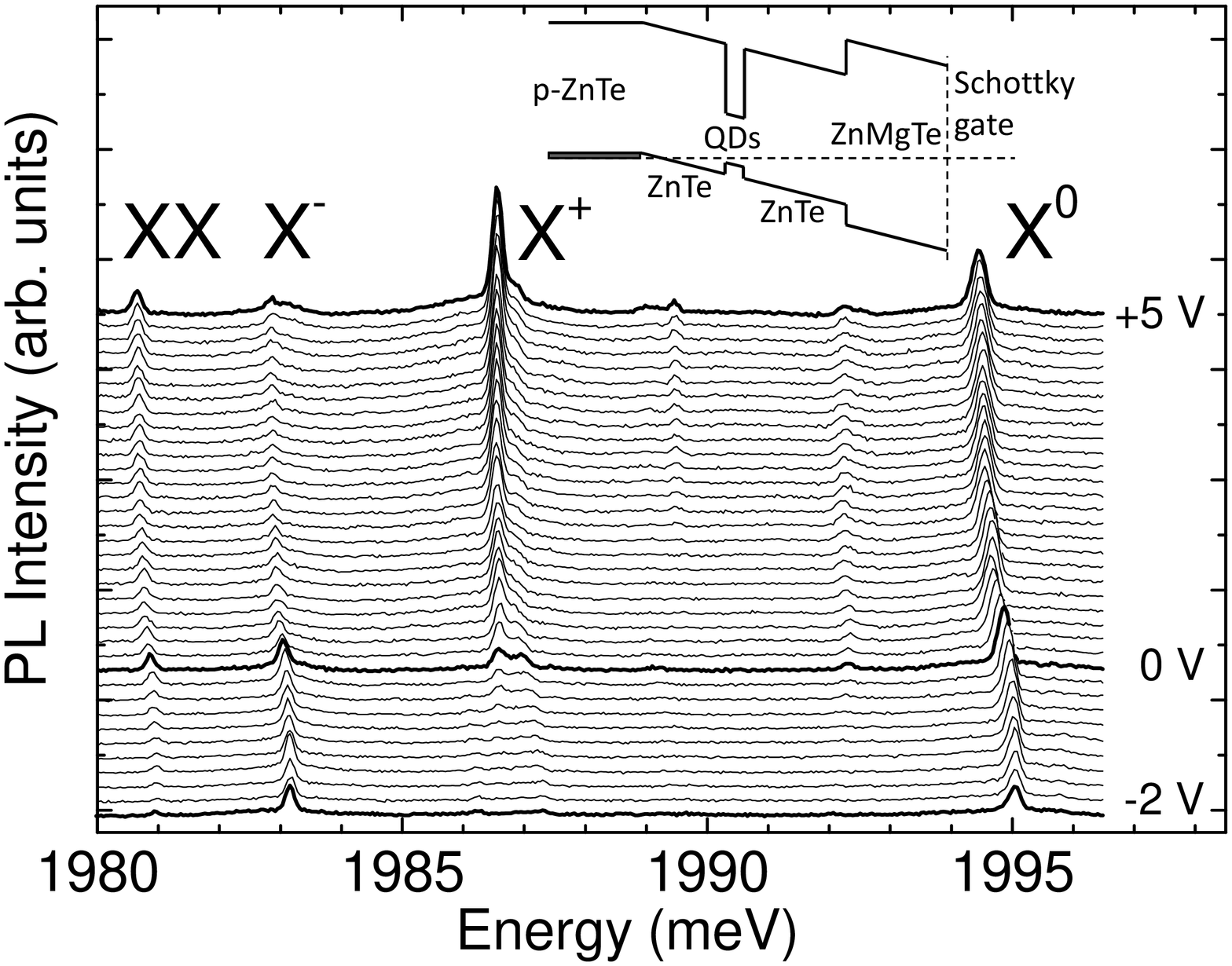}
  \caption{PL spectra of a single CdTe QD embedded in a field effect structure under bias: from $-2$ V on the bottom to $+5$ V on top. Thick line marks the spectrum at 0 V. Identified transitions, from low to high energies, correspond to biexciton, negatively and positively charged and neutral exciton recombinations. The inset shows the band structure under reverse bias.}
  \label{spectra}
\end{figure}

We start with discussing bias-dependent PL of a QD embedded in a Schottky diode structure. In Figure \ref{spectra}, we present spectra from the low energy tail of the ensemble PL band under applied bias: from reverse ($-2$ V, bottom spectrum) to forward ($+5$ V, top spectrum). We identify the observed transitions as recombinations of four excitonic complexes confined to the same single QD. Highest energy transition is the neutral exciton (X$^{0}$) recombination. At zero bias, positively (X$^{+}$) and negatively (X$^{-}$) charged excitons and a biexciton (XX) are red-shifted by 8.2, 11.8 and 14.9 meV, respectively. With increasing excitation power, we observe a roughly quadratic increase of the XX line and linear increase of the X$^{0}$ and charged exciton transitions.

The identification of the transition lines is based on the analysis of the intensity dependence on applied bias. Application of negative (reverse) bias (see inset in Fig. \ref{spectra}) enhances the tunneling of photocreated carriers out from the dots \cite{oul02,sei05}. Tunneling of holes is faster owing to a weak confinement in the valence band, which results from a vanishing valence band offset at the CdTe/ZnTe heterointerface. Therefore, at negative bias the X$^{-}$ dominates the spectrum. Applying an increasingly forward bias voltage restores the barrier between the back contact and the QD, and shifts the Fermi level above the QD states, thus injecting holes into the dots. As a result, at positive (forward) bias X$^{+}$ dominates the spectrum (see Fig. \ref{spectra}). In a charge tunable device, in order to inject a next carrier, the Fermi level needs to be lifted by a certain addition energy to suppress Coulomb blockade \cite{tar96,fra00}. As a result, distinct charging steps are usually  observed in the bias-dependent PL spectra\cite{war00,fin01,ebl06,edi07}. As seen in Fig. \ref{spectra}, in our results these steps are blurred and at a given bias voltage transitions related to various excitonic complexes are observed simultaneously. This coexistence of various charge states may result from a relatively weak tunnel coupling between the back contact and the dots \cite{bai01}. Moreover, single carriers can be captured from e.g. the wetting layer. Although we observe no emission from a wetting layer in our samples, the dots are reportedly formed on top of two dimensional platelets that provide a sequence of spatially extended excited states \cite{ngu07}. In the experiment, we excite the dot above these states enabling a stochastic capture of separate electrons and holes. The capture competes with bias-controlled charging and consequently charging steps are strongly masked and various charge states coexist.

The bias applied between the top Schottky gate and the p-type back contact generates an electric field along the growth axis. As a result, the transition lines are shifted due to the QCSE. For moderate electric fields, these shifts can be approximated by first two orders of the perturbation expansion: $E(F)=E(0)- p \cdot F + \beta \cdot F^2$, where $E(0)$ is the transition energy at zero electric field, and $p$ and $\beta$ are built-in dipole moment and electron-hole polarizability, respectively \cite{fin04}. In order to quantitatively analyze the Stark shifts, electric field magnitude $F$ has to be determined. In principle, $F$ is given by a simple capacitor formula: $F=(U-U_{bi})/d$, where $U$ and $U_{bi}$ are the applied and built-in voltage, respectively, and $d$ is the width of the intrinsic region. However, at a small forward bias the flow of charge screens out the external electric field. To analyze the shifts under such conditions, we fit the X$^{0}$ PL energy dependence on electric field $E_{X^0}(F)$ for the negative bias, and assume that X$^{0}$ transition at forward bias follows this dependence. In this way the voltage-to-electric field conversion is performed for the positive bias range. Another uncertainty in the determination of $F$ is related to a build-up of a space charge upon optical excitation. Photoexcited electrons can become trapped at the interface between the ZnTe capping layer and Zn$_{0.9}$Mg$_{0.1}$Te blocking barrier (see inset in Fig. \ref{spectra}), creating an electric field opposite in polarity to the applied bias -- an effect which was shown to affect the PL transition energies down to lowest excitation densities \cite{sei05}. In order to investigate Coulomb interactions in our dots, we chose to work at an excitation density, at which the biexciton transition is resolved. However, we checked that decreasing the power by a factor of 6 shifts the transition energies only by about 100 $\mu$eV, slightly more than our spectral resolution of 70 $\mu$eV and and much less than the spectral linewidth of $\sim 220 \mu$eV. Therefore, we conclude that the effects related to the build-up of the space charge are negligible.

\begin{figure}[!h]
  \includegraphics[angle=0,width=.5\textwidth]{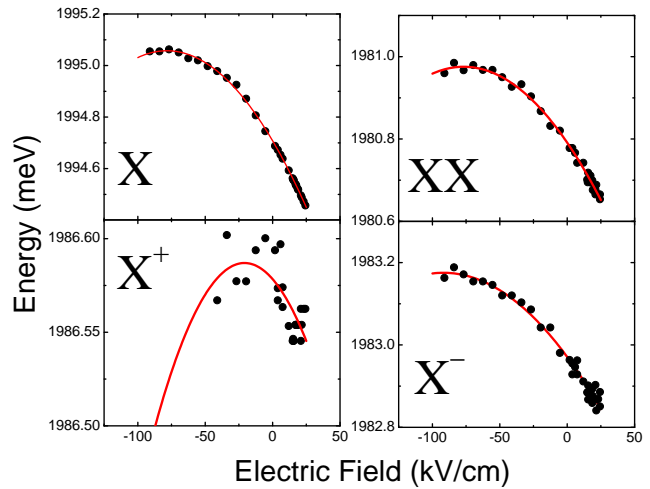}
  \caption{Transition energies as a function of the electric field for the four excitonic complexes. Lines are fitted second order polynomials.}
  \label{QCSE}
\end{figure}

Measured transition energies as a function of the applied electric field together with fitted curves are presented in Fig. \ref{QCSE}. Positive and negative $F$ values correspond to electric fields applied parallel and antiparallel to the growth axis, respectively. Clearly, the Stark shifts are correctly reproduced by a second order polynomial. From the fits in Fig. \ref{QCSE} we gain access to the built in dipole moment $p$ and the electron-hole polarizability $\beta$. The former corresponds to the zero field distance between the centers of gravity of electron and hole wave functions. In the case of X$^{0}$, we obtain $p/e=0.89\pm 0.03$ \AA\ and $\beta=5.7 \pm 0.89$ nm$^2$/V. Positive sign of $p$ indicates that in absence of electric field, the hole is located above the electron. This inverted carrier alignment points out a translational asymmetry along the growth axis \cite{fry00} possibly related to a Zn/Cd intermixing. As negative electric field is increased, the centers of electron and hole wavefunctions are brought together towards the dot center. Eventually, a cancelation of the dipole moment occurs at an electric field $F_0 \approx  -78 $kV$\cdot$cm$^{-1}$, where the transition energy dependence on $F$ reaches a maximum. The value of $\beta$ is of the same order of magnitude as in CdSe QDs \cite{seu01} and roughly an order of magnitude smaller than in InAs QDs \cite{fin04}. This latter results displays the influence of a stronger electron-hole attraction in II-VI nanostructures with respect to their less polar III-V counterparts.

\begin{table}[h]
%\begin{center}
\begin{tabularx}{7cm}{||X|X|X||}
  \hline
   & $p/e$ (\AA) & $\Delta p/p(\mbox{X}^0)$\\
  \hline
  X$^{0}$ & 0.89 $\pm$ 0.03 & ---  \\
  X$^{+}$ & 0.08 $\pm$ 0.02 & $-91$ \% \\
  X$^{-}$ & 0.45 $\pm$ 0.02 & $-49$ \% \\
  XX & 0.48 $\pm$ 0.01 & $-46$ \% \\
  \hline
\end{tabularx}
\caption{\label{bidm}Values of built-in dipole moment obtained for each of the excitonic complexes from fitting of the transition energies dependence on electric field and the its reduction relative to the X$^{0}$.}
%\end{center}
\vspace{-0.6cm}
\end{table}

Values of $p/e$ retrieved from the quadratic fits for all four investigated excitonic complexes are collected in Table \ref{bidm} together with the reduction $\Delta p$ relative to the X$^{0}$. We note that although the accuracy of the fit for the X$^+$ is lower than for the remaining complexes, it is clear that the X$^+$ energy shift in electric field is much smaller than for the other complexes. Therefore, from Table \ref{bidm} we infer that an addition of a second hole nearly cancels the built in dipole, while addition of a second electron reduces it only by roughly 50\%. This result implies that upon charging the redistribution of the hole wave function is more pronounced than the modification of the electron wave function.

\section{Time-resolved photoluminescence spectroscopy}

\begin{figure}[!b]
  \includegraphics[angle=0,width=.45\textwidth]{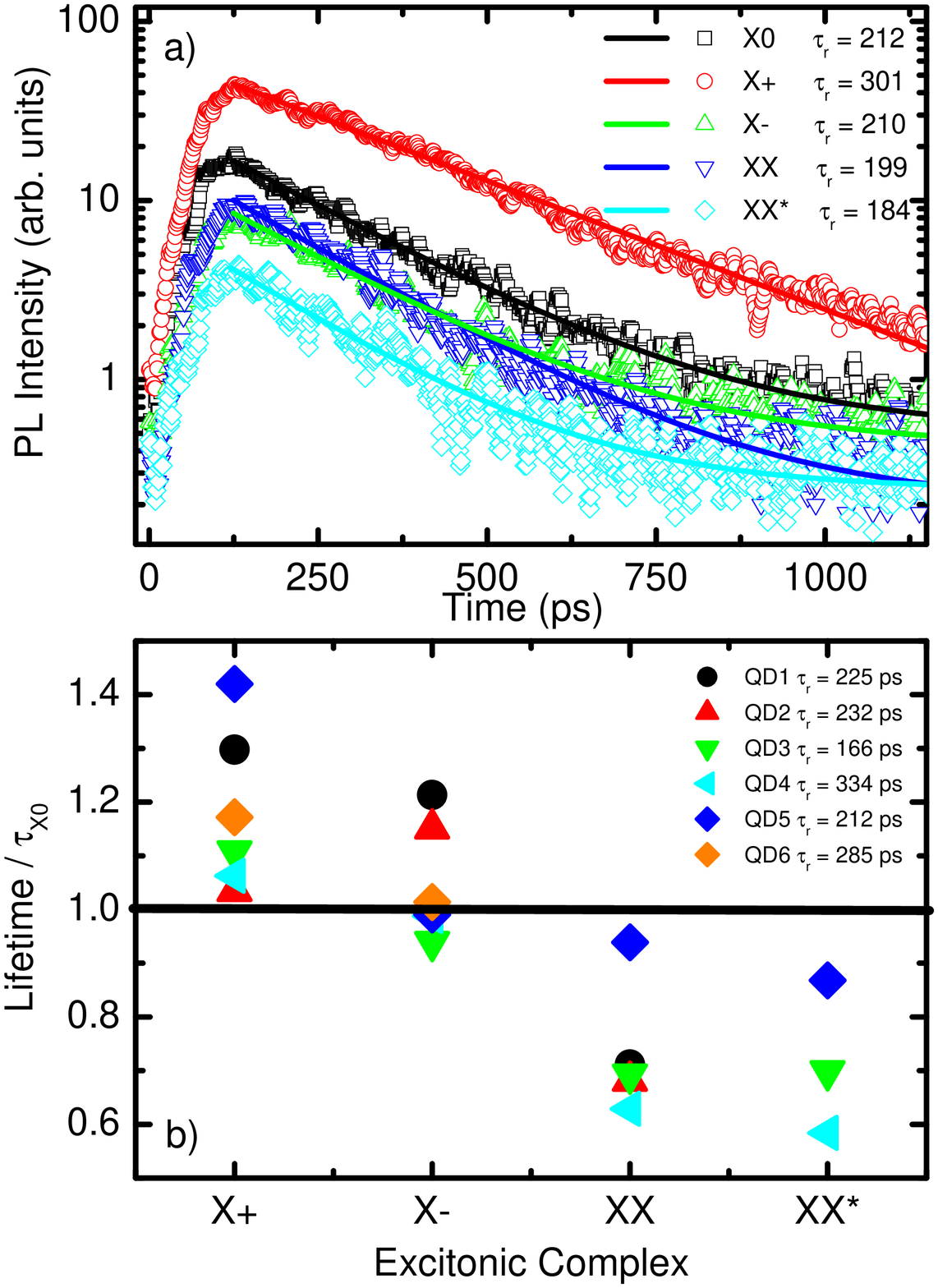}
  \caption{a) PL time traces of various excitonic complexes from a single quantum dot. Lines are single exponential decays with lifetimes of each complex given in the legend. b) Lifetimes of the observed excitonic complexes collected on six different dots, normalized to neutral exciton lifetime. }
  \label{lifetimes}
\end{figure}

In order to gain more insight into the modifications of carrier wave functions upon charging, we performed measurements of recombinations lifetimes $\tau$. In Fig. \ref{lifetimes}a, we present PL transients from a single QD, collected for the X$^{0}$, X$^{+}$, X$^{-}$, XX, and another transition, labeled as XX$^{*}$, which appears redshifted with respect to the XX and has a similar power dependence. Most probably it is a recombination of a negatively charged biexciton \cite{kaz09}. All transients exhibit small slow component, treated as a background, related to carrier recapture or dark exciton recombination. A single exponential decay was fitted to extract the lifetimes and the numerical results are presented in the legend. In Fig. \ref{lifetimes}b, we collect the lifetimes obtained from six different dots. To address a relative modification of carrier wave function upon charging, we present the lifetimes in units of the X$^{0}$ lifetime $\tau(X^0)$ , given in the legend. We find that upon charging with an extra hole, the recombination lifetime is notably increased -- on average $\tau(X^+)/ \tau(X^0) = 1.18 \pm 0.06$. On the other hand, charging with an extra electron affects the lifetime only slightly -- averaging over our data we find $\tau(X^-)/ \tau(X^0)= 1.05 \pm 0.04$.
The biexciton and XX$^{*}$ lifetimes are shortened -- $\tau(XX)/\tau(X^0) = 0.73 \pm 0.05 \tau(X^0)$ and $\tau(XX^*)/\tau(X^0) = 0.72 \pm 0.08 \tau(X^0)$.

These results again point out the relative stiffness and softness of the electron and hole wave functions, respectively. Indeed, the increase of the X$^{+}$ lifetime together with almost complete cancelation of the built-in dipole (see Table \ref{bidm}) indicate that upon addition of an extra hole, its wave function expands laterally decreasing both the electron-hole overlap and the dipole. On the other hand, addition of an extra electron does not result in an important change of the electron wave function as both the X$^-$ built-in dipole and the lifetime remain on average only weakly affected. The redistribution of carrier wave functions upon charging are most clearly demonstrated in the behavior of the XX decay. Neglecting the wave function modifications upon changing of the dot occupation, one expects the XX lifetime to be exactly 0.5 $ \cdot \tau(X^0)$ since the biexciton has two decay channels, while X$^0$ has only one \cite{feu08}. Instead, we only observe a decrease by a factor of 0.73, which directly reflects the decrease of the electron-hole overlap. We suppose that the decrease stems mostly from the presence of a second hole.

The lifetime data allows us to address also the confinement conditions in our QDs. In the strong confinement limit, carrier wave functions are determined by the QD potential and Coulomb interactions are only a small perturbation. In such a case, the wave functions are stiff and changing the dot occupancy modifies them only slightly. As a consequence, the transition lifetimes do not depend on the dot charge state and within a two level model are given by \cite{lou73}:

\be
\tau = \frac{3 \lambda_{PL}^2 \epsilon_0 c m_0}{2 \pi n e^2 f}
\label{tau}
\ee
where $\lambda_{PL}$ is the PL wavelength, $n$ is the refractive index of the medium surrounding the QD and $f$ -- the oscillator strength proportional to the overlap integral $\langle \phi_e|\phi_h \rangle$:\cite{kno63}

\be
f=\langle \phi_e|\phi_h \rangle \frac{E_P}{2 E_{PL}}
\label{ef}
\ee
where $E_{PL}$ is the PL photon energy and $E_{P}$ is the Kane energy, 17.9 eV for CdTe. In the strong confinement limit $\langle \phi_e|\phi_h \rangle=1$ and the above formula yields for a CdTe QD in a ZnTe matrix ($n=3.0$) emitting in 2.0 eV a lifetime of 1.3 ns. This is much longer than experimentally observed lifetimes on the order of 300 ps. Moreover, as evidenced in Fig. \ref{lifetimes}, the lifetimes clearly depend on the QD charge state. We can therefore conclude that our system is far from the strong confinement limit. Indeed, bulk exciton Bohr radius for CdTe is 3 nm, about ten times smaller than the dot lateral size. Moreover, as pointed out above, the confinement in the valence band is particularly weak owing to a vanishing valence band offset and it results mainly from strain. On the other hand, confinement in the conduction band is stronger, since all the band mismatch between CdTe and ZnTe gives rise to a potential well for electrons.

\section{Discussion and Conclusions}

The analysis of the reduction of the built-in dipole moment (see Table I) and the lifetime dependence on the dot occupation lead us to a conclusion that the hole wave function is soft, while the electron wave function is rigid. The former undergoes a redistribution upon charging a dot, while the latter remains almost unaffected. The relative softness and stiffness of the hole and electron wave functions, respectively point to a stronger correlation among holes than electrons. The strong hole correlations are related to relatively weak confinement in the valence band, which provides closely spaced shells. In the language of configuration mixing it implies that the ground state holes admix easily other configurations, while electron wave function is predominantly built from the single particle s-shell orbital since higher lying shells are well separated in energy.\cite{bry88,cor03} We remark that the same conclusions were drawn for InGaAs QDs despite an entirely different potential depths and dielectric constants \cite{dal08,cli08}.

The regime of the weak confinement and strong Coulomb correlations in the valence band are also manifested in absolute values of the exciton lifetimes. In the strong confinement limit, we expect the lifetimes about four times longer than the observed ones. Under weak confinement, the lifetime of an excitonic complex cannot be described by a simple overlap integral as in Eqs. \ref{tau} and \ref{ef}, since the wave function is now a correlated one, e.g.:\cite{bry88}

\be
\Psi(\text{\bf{r}}_e,\text{\bf{r}}_h) = \phi_e(\text{\bf{r}}_e) \phi_h(\text{\bf{r}}_h) s(\rho_e-\rho_h)
\ee
where $\rho_{e,h}$ are in-plane electron and hole coordinates. The recombination rate is proportional to the probability of finding the electron and the hole at the same location, i.e. to $s(0)$ and is expected to increase with increasing QD radius \cite{bel98,cor03,hou05}. The weak confinement is also manifested by a clear dependence of the lifetime on the dot charge state -- an effect which would be absent, if the wave functions were frozen and given by single orbitals determined solely by the confinement.

\begin{figure}[!h]
  \includegraphics[angle=-90,width=.45\textwidth]{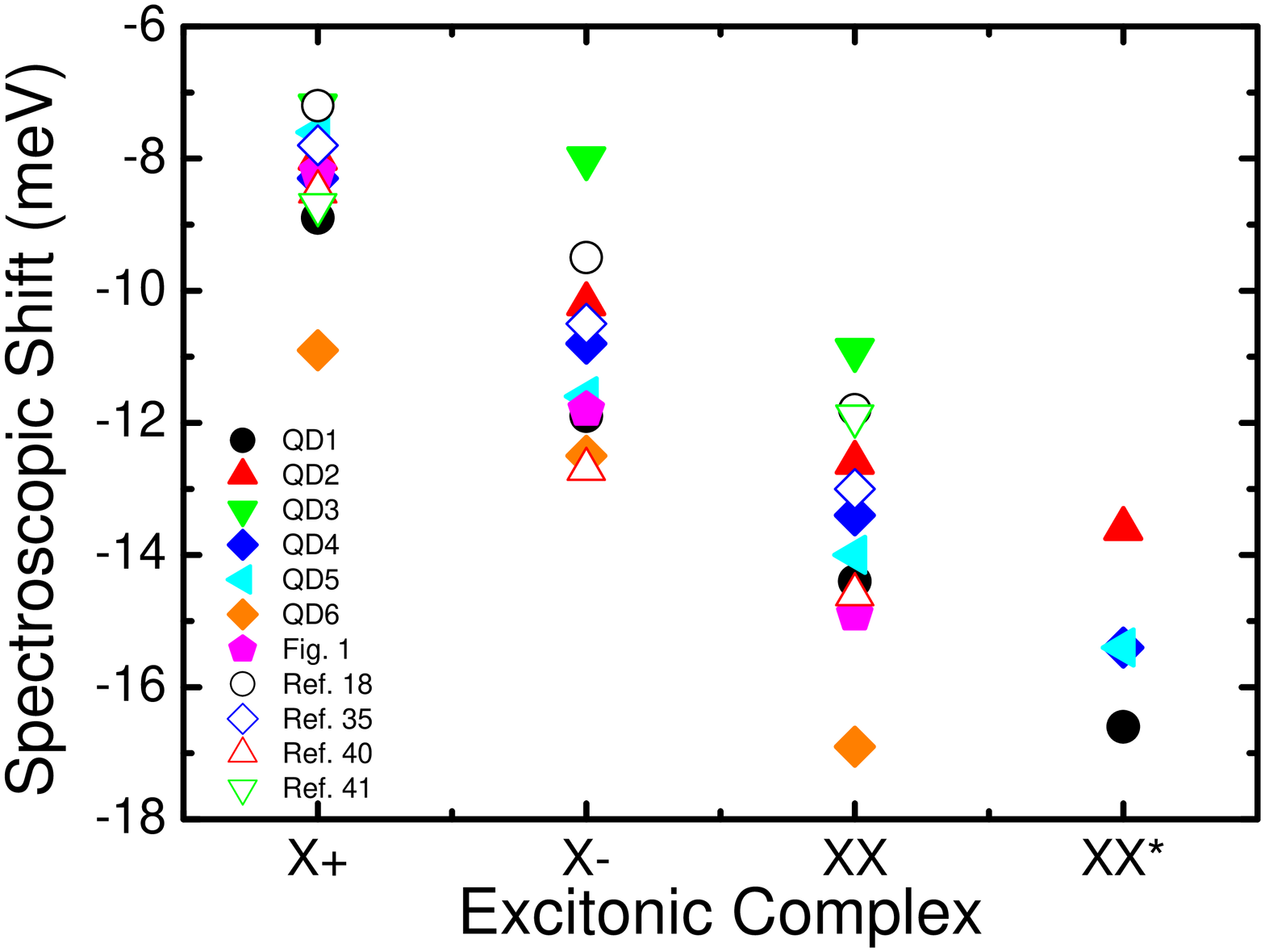}
  \caption{Spectroscopic shifts for X$^+$, X$^-$, XX, and XX$^*$ resolved in the present study and compared to values found in literature. In all cases, these excitonic complexes appear redshifted with respect to the X$^0$ and the same sequence of transitions is found.}
  \label{specshifts}
\end{figure}

The dominant role of correlations is further demonstrated in the shape of the emission spectrum of a single dot. We find the same sequence of the transition lines in all the dots studied {\em and} found in literature. In Figure \ref{specshifts}, we present the spectroscopic shifts of the observed excitonic complexes taken as the difference between a given complex and a neutral exciton X$^0$. We collect the data from the samples used in the present study -- the field effect sample discussed in Sec. III, and the samples used in the time-resolved PL studies in Sec. IV. We note that in {\em all} the studied dots, the charged excitons and biexcitons appear {\em redshifted} with respect to the X$^0$. Moreover, for all the studied dots, we record the same transition sequence: $E_{X^0}>E_{X^+}>E_{X^-}>E_{XX}>E_{XX^*}$. We also note that the same sequence was observed in all other CdTe QDs reported previously, where the charge state of the emitting complex was identified\cite{leg06,suf06,kow06b,kaz09}.

The redshift of all the charged exciton and biexciton transitions points out that the direct Coulomb interactions are weaker than correlations as a result of relatively weak confinement. Consequently, the emission spectrum qualitatively resembles the one observed in 2D systems, where the charged exciton binding energies are also dominated by correlation effects \cite{riv00}. This remains in stark contrast with InGaAs system, where charged exciton transitions may appear on both sides of X$^0$ depending on the localization degree of electrons and holes, which is controlled by the dot morphology \cite{fin01,bes03,fin04,cli08,dal08}.

In conclusion, we investigated the impact of Coulomb interactions on carrier wave functions in single CdTe quantum dots by means of Stark spectroscopy and time-resolved photoluminescence. We found that in absence of electric field, the hole is located above the electron giving rise to an electric dipole. Moreover, we discovered that
the hole wave function undergoes strong modifications upon changing of the dot charge state, while the electron wave function is nearly unaffected. We attribute this effect to a relatively weak confinement in the valence band, which makes Coulomb correlations dominate over the confinement. As a result, the emission spectrum of single dots exhibit the same sequence of transitions related to different charge states with the neutral exciton emitting at highest energy as in 2D systems.

%\begin{theacknowledgments}
We would like to thank M. Korkusiński, P. Hawrylak and Ł. Cywiński for fruitful discussions. This research was supported by a Polish Ministry of Science and Education grant no. 0634/BH03/2007/33 and the Polonium Programme and by European Union within European Regional Development Fund, through Innovative Economy grant (POIG.01.01.02-00-008/08) and SANDiE Network of Excellence.
%\end{theacknowledgments}

%}


\begin{thebibliography}{39}%
\makeatletter
\providecommand \@ifxundefined [1]{%
 \@ifx{#1\undefined}
}%
\providecommand \@ifnum [1]{%
 \ifnum #1\expandafter \@firstoftwo
 \else \expandafter \@secondoftwo
 \fi
}%
\providecommand \@ifx [1]{%
 \ifx #1\expandafter \@firstoftwo
 \else \expandafter \@secondoftwo
 \fi
}%
\providecommand \natexlab [1]{#1}%
\providecommand \enquote  [1]{``#1''}%
\providecommand \bibnamefont  [1]{#1}%
\providecommand \bibfnamefont [1]{#1}%
\providecommand \citenamefont [1]{#1}%
\providecommand \href@noop [0]{\@secondoftwo}%
\providecommand \href [0]{\begingroup \@sanitize@url \@href}%
\providecommand \@href[1]{\@@startlink{#1}\@@href}%
\providecommand \@@href[1]{\endgroup#1\@@endlink}%
\providecommand \@sanitize@url [0]{\catcode `\\12\catcode `\$12\catcode
  `\&12\catcode `\#12\catcode `\^12\catcode `\_12\catcode `\%12\relax}%
\providecommand \@@startlink[1]{}%
\providecommand \@@endlink[0]{}%
\providecommand \url  [0]{\begingroup\@sanitize@url \@url }%
\providecommand \@url [1]{\endgroup\@href {#1}{\urlprefix }}%
\providecommand \urlprefix  [0]{URL }%
\providecommand \Eprint [0]{\href }%
\@ifxundefined \urlstyle {%
  \providecommand \doi  [0]{\begingroup \@sanitize@url \@doi}%
  \providecommand \@doi [1]{\endgroup \@@startlink {\doibase
  #1}doi:\discretionary {}{}{}#1\@@endlink }%
}{%
  \providecommand \doi  [0]{doi:\discretionary{}{}{}\begingroup
  \urlstyle{rm}\Url }%
}%
\providecommand \doibase [0]{http://dx.doi.org/}%
\providecommand \Doi [0]{\begingroup \@sanitize@url \@Doi }%
\providecommand \@Doi  [1]{\endgroup\@@startlink{\doibase#1}\@@Doi}%
\providecommand \@@Doi [1]{#1\@@endlink}%
\providecommand \selectlanguage [0]{\@gobble}%
\providecommand \bibinfo  [0]{\@secondoftwo}%
\providecommand \bibfield  [0]{\@secondoftwo}%
\providecommand \translation [1]{[#1]}%
\providecommand \BibitemOpen [0]{}%
\providecommand \bibitemStop [0]{}%
\providecommand \bibitemNoStop [0]{.\EOS\space}%
\providecommand \EOS [0]{\spacefactor3000\relax}%
\providecommand \BibitemShut  [1]{\csname bibitem#1\endcsname}%
%</preamble>
\bibitem [{\citenamefont {Drexler}\ \emph {et~al.}(1994)\citenamefont
  {Drexler}, \citenamefont {Leonard}, \citenamefont {Hansen}, \citenamefont
  {Kotthaus},\ and\ \citenamefont {Petroff}}]{dre94}%
  \BibitemOpen
  \bibfield  {author} {\bibinfo {author} {\bibfnamefont {H.}~\bibnamefont
  {Drexler}}, \bibinfo {author} {\bibfnamefont {D.}~\bibnamefont {Leonard}},
  \bibinfo {author} {\bibfnamefont {W.}~\bibnamefont {Hansen}}, \bibinfo
  {author} {\bibfnamefont {J.~P.}\ \bibnamefont {Kotthaus}}, \ and\ \bibinfo
  {author} {\bibfnamefont {P.~M.}\ \bibnamefont {Petroff}},\ }\href@noop {}
  {\bibfield  {journal} {\bibinfo  {journal} {Phys. Rev. Lett.},\ }\textbf
  {\bibinfo {volume} {73}},\ \bibinfo {pages} {2252} (\bibinfo {year}
  {1994})}\BibitemShut {NoStop}%
\bibitem [{\citenamefont {Regelman}\ \emph {et~al.}(2001)\citenamefont
  {Regelman}, \citenamefont {Dekel}, \citenamefont {Gershoni}, \citenamefont
  {Ehrenfreund}, \citenamefont {Williamson}, \citenamefont {Shumway},
  \citenamefont {Zunger}, \citenamefont {Schoenfeld},\ and\ \citenamefont
  {Petroff}}]{reg01}%
  \BibitemOpen
  \bibfield  {author} {\bibinfo {author} {\bibfnamefont {D.~V.}\ \bibnamefont
  {Regelman}}, \bibinfo {author} {\bibfnamefont {E.}~\bibnamefont {Dekel}},
  \bibinfo {author} {\bibfnamefont {D.}~\bibnamefont {Gershoni}}, \bibinfo
  {author} {\bibfnamefont {E.}~\bibnamefont {Ehrenfreund}}, \bibinfo {author}
  {\bibfnamefont {A.~J.}\ \bibnamefont {Williamson}}, \bibinfo {author}
  {\bibfnamefont {J.}~\bibnamefont {Shumway}}, \bibinfo {author} {\bibfnamefont
  {A.}~\bibnamefont {Zunger}}, \bibinfo {author} {\bibfnamefont {W.~V.}\
  \bibnamefont {Schoenfeld}}, \ and\ \bibinfo {author} {\bibfnamefont {P.~M.}\
  \bibnamefont {Petroff}},\ }\Doi {10.1103/PhysRevB.64.165301} {\bibfield
  {journal} {\bibinfo  {journal} {Phys. Rev. B},\ }\textbf {\bibinfo {volume}
  {64}},\ \bibinfo {pages} {165301} (\bibinfo {year} {2001})}\BibitemShut
  {NoStop}%
\bibitem [{\citenamefont {Bester}\ and\ \citenamefont {Zunger}(2003)}]{bes03}%
  \BibitemOpen
  \bibfield  {author} {\bibinfo {author} {\bibfnamefont {G.}~\bibnamefont
  {Bester}}\ and\ \bibinfo {author} {\bibfnamefont {A.}~\bibnamefont
  {Zunger}},\ }\Doi {10.1103/PhysRevB.68.073309} {\bibfield  {journal}
  {\bibinfo  {journal} {Phys. Rev. B},\ }\textbf {\bibinfo {volume} {68}},\
  \bibinfo {pages} {073309} (\bibinfo {year} {2003})}\BibitemShut {NoStop}%
\bibitem [{\citenamefont {Dalgarno}\ \emph {et~al.}(2008)\citenamefont
  {Dalgarno}, \citenamefont {Smith}, \citenamefont {McFarlane}, \citenamefont
  {Gerardot}, \citenamefont {Karrai}, \citenamefont {Badolato}, \citenamefont
  {Petroff},\ and\ \citenamefont {Warburton}}]{dal08}%
  \BibitemOpen
  \bibfield  {author} {\bibinfo {author} {\bibfnamefont {P.~A.}\ \bibnamefont
  {Dalgarno}}, \bibinfo {author} {\bibfnamefont {J.~M.}\ \bibnamefont {Smith}},
  \bibinfo {author} {\bibfnamefont {J.}~\bibnamefont {McFarlane}}, \bibinfo
  {author} {\bibfnamefont {B.~D.}\ \bibnamefont {Gerardot}}, \bibinfo {author}
  {\bibfnamefont {K.}~\bibnamefont {Karrai}}, \bibinfo {author} {\bibfnamefont
  {A.}~\bibnamefont {Badolato}}, \bibinfo {author} {\bibfnamefont {P.~M.}\
  \bibnamefont {Petroff}}, \ and\ \bibinfo {author} {\bibfnamefont {R.~J.}\
  \bibnamefont {Warburton}},\ }\Doi {10.1103/PhysRevB.77.245311} {\bibfield
  {journal} {\bibinfo  {journal} {Phys. Rev. B},\ }\textbf {\bibinfo {volume}
  {77}},\ \bibinfo {eid} {245311} (\bibinfo {year} {2008})}\BibitemShut
  {NoStop}%
\bibitem [{\citenamefont {Climente}\ \emph {et~al.}(2008)\citenamefont
  {Climente}, \citenamefont {Bertoni},\ and\ \citenamefont {Goldoni}}]{cli08}%
  \BibitemOpen
  \bibfield  {author} {\bibinfo {author} {\bibfnamefont {J.~I.}\ \bibnamefont
  {Climente}}, \bibinfo {author} {\bibfnamefont {A.}~\bibnamefont {Bertoni}}, \
  and\ \bibinfo {author} {\bibfnamefont {G.}~\bibnamefont {Goldoni}},\ }\Doi
  {10.1103/PhysRevB.78.155316} {\bibfield  {journal} {\bibinfo  {journal}
  {Phys. Rev. B},\ }\textbf {\bibinfo {volume} {78}},\ \bibinfo {eid} {155316}
  (\bibinfo {year} {2008})}\BibitemShut {NoStop}%
\bibitem [{\citenamefont {Tarucha}\ \emph {et~al.}(1996)\citenamefont
  {Tarucha}, \citenamefont {Austing}, \citenamefont {Honda}, \citenamefont
  {van~der Hage},\ and\ \citenamefont {Kouwenhoven}}]{tar96}%
  \BibitemOpen
  \bibfield  {author} {\bibinfo {author} {\bibfnamefont {S.}~\bibnamefont
  {Tarucha}}, \bibinfo {author} {\bibfnamefont {D.~G.}\ \bibnamefont
  {Austing}}, \bibinfo {author} {\bibfnamefont {T.}~\bibnamefont {Honda}},
  \bibinfo {author} {\bibfnamefont {R.~J.}\ \bibnamefont {van~der Hage}}, \
  and\ \bibinfo {author} {\bibfnamefont {L.~P.}\ \bibnamefont {Kouwenhoven}},\
  }\Doi {10.1103/PhysRevLett.77.3613} {\bibfield  {journal} {\bibinfo
  {journal} {Phys. Rev. Lett.},\ }\textbf {\bibinfo {volume} {77}},\ \bibinfo
  {pages} {3613} (\bibinfo {year} {1996})}\BibitemShut {NoStop}%
\bibitem [{\citenamefont {Warburton}\ \emph {et~al.}(2000)\citenamefont
  {Warburton}, \citenamefont {Sch\"{a}flein}, \citenamefont {Haft},
  \citenamefont {Bickel}, \citenamefont {Lorke}, \citenamefont {Karrai},
  \citenamefont {Garcia}, \citenamefont {Schoenfeld},\ and\ \citenamefont
  {Petroff}}]{war00}%
  \BibitemOpen
  \bibfield  {author} {\bibinfo {author} {\bibfnamefont {R.~J.}\ \bibnamefont
  {Warburton}}, \bibinfo {author} {\bibfnamefont {C.}~\bibnamefont
  {Sch\"{a}flein}}, \bibinfo {author} {\bibfnamefont {D.}~\bibnamefont {Haft}},
  \bibinfo {author} {\bibfnamefont {F.}~\bibnamefont {Bickel}}, \bibinfo
  {author} {\bibfnamefont {A.}~\bibnamefont {Lorke}}, \bibinfo {author}
  {\bibfnamefont {K.}~\bibnamefont {Karrai}}, \bibinfo {author} {\bibfnamefont
  {J.~M.}\ \bibnamefont {Garcia}}, \bibinfo {author} {\bibfnamefont
  {W.}~\bibnamefont {Schoenfeld}}, \ and\ \bibinfo {author} {\bibfnamefont
  {P.~M.}\ \bibnamefont {Petroff}},\ }\href@noop {} {\bibfield  {journal}
  {\bibinfo  {journal} {Nature},\ }\textbf {\bibinfo {volume} {405}},\ \bibinfo
  {pages} {926} (\bibinfo {year} {2000})}\BibitemShut {NoStop}%
\bibitem [{\citenamefont {Finley}\ \emph {et~al.}(2001)\citenamefont {Finley},
  \citenamefont {Fry}, \citenamefont {Ashmore}, \citenamefont {Lema\^itre},
  \citenamefont {Tartakovskii}, \citenamefont {Oulton}, \citenamefont
  {Mowbray}, \citenamefont {Skolnick}, \citenamefont {Hopkinson}, \citenamefont
  {Buckle},\ and\ \citenamefont {Maksym}}]{fin01}%
  \BibitemOpen
  \bibfield  {author} {\bibinfo {author} {\bibfnamefont {J.~J.}\ \bibnamefont
  {Finley}}, \bibinfo {author} {\bibfnamefont {P.~W.}\ \bibnamefont {Fry}},
  \bibinfo {author} {\bibfnamefont {A.~D.}\ \bibnamefont {Ashmore}}, \bibinfo
  {author} {\bibfnamefont {A.}~\bibnamefont {Lema\^itre}}, \bibinfo {author}
  {\bibfnamefont {A.~I.}\ \bibnamefont {Tartakovskii}}, \bibinfo {author}
  {\bibfnamefont {R.}~\bibnamefont {Oulton}}, \bibinfo {author} {\bibfnamefont
  {D.~J.}\ \bibnamefont {Mowbray}}, \bibinfo {author} {\bibfnamefont {M.~S.}\
  \bibnamefont {Skolnick}}, \bibinfo {author} {\bibfnamefont {M.}~\bibnamefont
  {Hopkinson}}, \bibinfo {author} {\bibfnamefont {P.~D.}\ \bibnamefont
  {Buckle}}, \ and\ \bibinfo {author} {\bibfnamefont {P.~A.}\ \bibnamefont
  {Maksym}},\ }\Doi {10.1103/PhysRevB.63.161305} {\bibfield  {journal}
  {\bibinfo  {journal} {Phys. Rev. B},\ }\textbf {\bibinfo {volume} {63}},\
  \bibinfo {pages} {161305} (\bibinfo {year} {2001})}\BibitemShut {NoStop}%
\bibitem [{\citenamefont {Baier}\ \emph {et~al.}(2001)\citenamefont {Baier},
  \citenamefont {Findeis}, \citenamefont {Zrenner}, \citenamefont {Bichler},\
  and\ \citenamefont {Abstreiter}}]{bai01}%
  \BibitemOpen
  \bibfield  {author} {\bibinfo {author} {\bibfnamefont {M.}~\bibnamefont
  {Baier}}, \bibinfo {author} {\bibfnamefont {F.}~\bibnamefont {Findeis}},
  \bibinfo {author} {\bibfnamefont {A.}~\bibnamefont {Zrenner}}, \bibinfo
  {author} {\bibfnamefont {M.}~\bibnamefont {Bichler}}, \ and\ \bibinfo
  {author} {\bibfnamefont {G.}~\bibnamefont {Abstreiter}},\ }\Doi
  {10.1103/PhysRevB.64.195326} {\bibfield  {journal} {\bibinfo  {journal}
  {Phys. Rev. B},\ }\textbf {\bibinfo {volume} {64}},\ \bibinfo {pages}
  {195326} (\bibinfo {year} {2001})}\BibitemShut {NoStop}%
\bibitem [{\citenamefont {Ediger}\ \emph {et~al.}(2007)\citenamefont {Ediger},
  \citenamefont {Bester}, \citenamefont {Badolato}, \citenamefont {Petroff},
  \citenamefont {Karrai}, \citenamefont {Zunger},\ and\ \citenamefont
  {Warburton}}]{edi07}%
  \BibitemOpen
  \bibfield  {author} {\bibinfo {author} {\bibfnamefont {M.}~\bibnamefont
  {Ediger}}, \bibinfo {author} {\bibfnamefont {G.}~\bibnamefont {Bester}},
  \bibinfo {author} {\bibfnamefont {A.}~\bibnamefont {Badolato}}, \bibinfo
  {author} {\bibfnamefont {P.~M.}\ \bibnamefont {Petroff}}, \bibinfo {author}
  {\bibfnamefont {K.}~\bibnamefont {Karrai}}, \bibinfo {author} {\bibfnamefont
  {A.}~\bibnamefont {Zunger}}, \ and\ \bibinfo {author} {\bibfnamefont {R.~J.}\
  \bibnamefont {Warburton}},\ }\href@noop {} {\bibfield  {journal} {\bibinfo
  {journal} {Nature Physics},\ }\textbf {\bibinfo {volume} {3}},\ \bibinfo
  {pages} {774} (\bibinfo {year} {2007})}\BibitemShut {NoStop}%
\bibitem [{\citenamefont {Eble}\ \emph {et~al.}(2006)\citenamefont {Eble},
  \citenamefont {Krebs}, \citenamefont {Lemaitre}, \citenamefont {Kowalik},
  \citenamefont {Kudelski}, \citenamefont {Voisin}, \citenamefont {Urbaszek},
  \citenamefont {Marie},\ and\ \citenamefont {Amand}}]{ebl06}%
  \BibitemOpen
  \bibfield  {author} {\bibinfo {author} {\bibfnamefont {B.}~\bibnamefont
  {Eble}}, \bibinfo {author} {\bibfnamefont {O.}~\bibnamefont {Krebs}},
  \bibinfo {author} {\bibfnamefont {A.}~\bibnamefont {Lemaitre}}, \bibinfo
  {author} {\bibfnamefont {K.}~\bibnamefont {Kowalik}}, \bibinfo {author}
  {\bibfnamefont {A.}~\bibnamefont {Kudelski}}, \bibinfo {author}
  {\bibfnamefont {P.}~\bibnamefont {Voisin}}, \bibinfo {author} {\bibfnamefont
  {B.}~\bibnamefont {Urbaszek}}, \bibinfo {author} {\bibfnamefont
  {X.}~\bibnamefont {Marie}}, \ and\ \bibinfo {author} {\bibfnamefont
  {T.}~\bibnamefont {Amand}},\ }\Doi {10.1103/PhysRevB.74.081306} {\bibfield
  {journal} {\bibinfo  {journal} {Phys. Rev. B},\ }\textbf {\bibinfo {volume}
  {74}},\ \bibinfo {eid} {081306} (\bibinfo {year} {2006})}\BibitemShut
  {NoStop}%
\bibitem [{\citenamefont {Tartakovskii}\ \emph {et~al.}(2007)\citenamefont
  {Tartakovskii}, \citenamefont {Wright}, \citenamefont {Russell},
  \citenamefont {Fal'ko}, \citenamefont {Van'kov}, \citenamefont
  {Skiba-Szymanska}, \citenamefont {Drouzas}, \citenamefont {Kolodka},
  \citenamefont {Skolnick}, \citenamefont {Fry}, \citenamefont {Tahraoui},
  \citenamefont {Liu},\ and\ \citenamefont {Hopkinson}}]{tar07}%
  \BibitemOpen
  \bibfield  {author} {\bibinfo {author} {\bibfnamefont {A.~I.}\ \bibnamefont
  {Tartakovskii}}, \bibinfo {author} {\bibfnamefont {T.}~\bibnamefont
  {Wright}}, \bibinfo {author} {\bibfnamefont {A.}~\bibnamefont {Russell}},
  \bibinfo {author} {\bibfnamefont {V.~I.}\ \bibnamefont {Fal'ko}}, \bibinfo
  {author} {\bibfnamefont {A.~B.}\ \bibnamefont {Van'kov}}, \bibinfo {author}
  {\bibfnamefont {J.}~\bibnamefont {Skiba-Szymanska}}, \bibinfo {author}
  {\bibfnamefont {I.}~\bibnamefont {Drouzas}}, \bibinfo {author} {\bibfnamefont
  {R.~S.}\ \bibnamefont {Kolodka}}, \bibinfo {author} {\bibfnamefont {M.~S.}\
  \bibnamefont {Skolnick}}, \bibinfo {author} {\bibfnamefont {P.~W.}\
  \bibnamefont {Fry}}, \bibinfo {author} {\bibfnamefont {A.}~\bibnamefont
  {Tahraoui}}, \bibinfo {author} {\bibfnamefont {H.-Y.}\ \bibnamefont {Liu}}, \
  and\ \bibinfo {author} {\bibfnamefont {M.}~\bibnamefont {Hopkinson}},\ }\Doi
  {10.1103/PhysRevLett.98.026806} {\bibfield  {journal} {\bibinfo  {journal}
  {Phys. Rev. Lett.},\ }\textbf {\bibinfo {volume} {98}},\ \bibinfo {eid}
  {026806} (\bibinfo {year} {2007})}\BibitemShut {NoStop}%
\bibitem [{\citenamefont {Ramsay}\ \emph {et~al.}(2008)\citenamefont {Ramsay},
  \citenamefont {Boyle}, \citenamefont {Kolodka}, \citenamefont {Oliveira},
  \citenamefont {Skiba-Szymanska}, \citenamefont {Liu}, \citenamefont
  {Hopkinson}, \citenamefont {Fox},\ and\ \citenamefont {Skolnick}}]{ram08}%
  \BibitemOpen
  \bibfield  {author} {\bibinfo {author} {\bibfnamefont {A.~J.}\ \bibnamefont
  {Ramsay}}, \bibinfo {author} {\bibfnamefont {S.~J.}\ \bibnamefont {Boyle}},
  \bibinfo {author} {\bibfnamefont {R.~S.}\ \bibnamefont {Kolodka}}, \bibinfo
  {author} {\bibfnamefont {J.~B.~B.}\ \bibnamefont {Oliveira}}, \bibinfo
  {author} {\bibfnamefont {J.}~\bibnamefont {Skiba-Szymanska}}, \bibinfo
  {author} {\bibfnamefont {H.~Y.}\ \bibnamefont {Liu}}, \bibinfo {author}
  {\bibfnamefont {M.}~\bibnamefont {Hopkinson}}, \bibinfo {author}
  {\bibfnamefont {A.~M.}\ \bibnamefont {Fox}}, \ and\ \bibinfo {author}
  {\bibfnamefont {M.~S.}\ \bibnamefont {Skolnick}},\ }\Doi
  {10.1103/PhysRevLett.100.197401} {\bibfield  {journal} {\bibinfo  {journal}
  {Phys. Rev. Lett.},\ }\textbf {\bibinfo {volume} {100}},\ \bibinfo {eid}
  {197401} (\bibinfo {year} {2008})}\BibitemShut {NoStop}%
\bibitem [{\citenamefont {L\'{e}ger}\ \emph {et~al.}(2006)\citenamefont
  {L\'{e}ger}, \citenamefont {Besombes}, \citenamefont {Fern\'{a}ndez-Rossier},
  \citenamefont {Maingault},\ and\ \citenamefont {Mariette}}]{leg06}%
  \BibitemOpen
  \bibfield  {author} {\bibinfo {author} {\bibfnamefont {Y.}~\bibnamefont
  {L\'{e}ger}}, \bibinfo {author} {\bibfnamefont {L.}~\bibnamefont {Besombes}},
  \bibinfo {author} {\bibfnamefont {J.}~\bibnamefont {Fern\'{a}ndez-Rossier}},
  \bibinfo {author} {\bibfnamefont {L.}~\bibnamefont {Maingault}}, \ and\
  \bibinfo {author} {\bibfnamefont {H.}~\bibnamefont {Mariette}},\ }\Doi
  {10.1103/PhysRevLett.97.107401} {\bibfield  {journal} {\bibinfo  {journal}
  {Phys. Rev. Lett.},\ }\textbf {\bibinfo {volume} {97}},\ \bibinfo {eid}
  {107401} (\bibinfo {year} {2006})}\BibitemShut {NoStop}%
\bibitem [{\citenamefont {Empedocles}\ and\ \citenamefont
  {Bawendi}(1997)}]{emp97}%
  \BibitemOpen
  \bibfield  {author} {\bibinfo {author} {\bibfnamefont {S.~A.}\ \bibnamefont
  {Empedocles}}\ and\ \bibinfo {author} {\bibfnamefont {M.~G.}\ \bibnamefont
  {Bawendi}},\ }\href@noop {} {\bibfield  {journal} {\bibinfo  {journal}
  {Science},\ }\textbf {\bibinfo {volume} {278}},\ \bibinfo {pages} {2114}
  (\bibinfo {year} {1997})}\BibitemShut {NoStop}%
\bibitem [{\citenamefont {Raymond}\ \emph {et~al.}(1998)\citenamefont
  {Raymond}, \citenamefont {Reynolds}, \citenamefont {Merz}, \citenamefont
  {Fafard}, \citenamefont {Feng},\ and\ \citenamefont {Charbonneau}}]{ray98}%
  \BibitemOpen
  \bibfield  {author} {\bibinfo {author} {\bibfnamefont {S.}~\bibnamefont
  {Raymond}}, \bibinfo {author} {\bibfnamefont {J.~P.}\ \bibnamefont
  {Reynolds}}, \bibinfo {author} {\bibfnamefont {J.~L.}\ \bibnamefont {Merz}},
  \bibinfo {author} {\bibfnamefont {S.}~\bibnamefont {Fafard}}, \bibinfo
  {author} {\bibfnamefont {Y.}~\bibnamefont {Feng}}, \ and\ \bibinfo {author}
  {\bibfnamefont {S.}~\bibnamefont {Charbonneau}},\ }\Doi
  {10.1103/PhysRevB.58.R13415} {\bibfield  {journal} {\bibinfo  {journal}
  {Phys. Rev. B},\ }\textbf {\bibinfo {volume} {58}},\ \bibinfo {pages}
  {R13415} (\bibinfo {year} {1998})}\BibitemShut {NoStop}%
\bibitem [{\citenamefont {Seufert}\ \emph {et~al.}(2001)\citenamefont
  {Seufert}, \citenamefont {Obert}, \citenamefont {Scheibner}, \citenamefont
  {Gippius}, \citenamefont {Bacher}, \citenamefont {Forchel}, \citenamefont
  {Passow}, \citenamefont {Leonardi},\ and\ \citenamefont {Hommel}}]{seu01}%
  \BibitemOpen
  \bibfield  {author} {\bibinfo {author} {\bibfnamefont {J.}~\bibnamefont
  {Seufert}}, \bibinfo {author} {\bibfnamefont {M.}~\bibnamefont {Obert}},
  \bibinfo {author} {\bibfnamefont {M.}~\bibnamefont {Scheibner}}, \bibinfo
  {author} {\bibfnamefont {N.~A.}\ \bibnamefont {Gippius}}, \bibinfo {author}
  {\bibfnamefont {G.}~\bibnamefont {Bacher}}, \bibinfo {author} {\bibfnamefont
  {A.}~\bibnamefont {Forchel}}, \bibinfo {author} {\bibfnamefont
  {T.}~\bibnamefont {Passow}}, \bibinfo {author} {\bibfnamefont
  {K.}~\bibnamefont {Leonardi}}, \ and\ \bibinfo {author} {\bibfnamefont
  {D.}~\bibnamefont {Hommel}},\ }\Doi {10.1063/1.1389504} {\bibfield  {journal}
  {\bibinfo  {journal} {App. Phys. Lett.},\ }\textbf {\bibinfo {volume} {79}},\
  \bibinfo {pages} {1033} (\bibinfo {year} {2001})}\BibitemShut {NoStop}%
\bibitem [{\citenamefont {Finley}\ \emph {et~al.}(2004)\citenamefont {Finley},
  \citenamefont {Sabathil}, \citenamefont {Vogl}, \citenamefont {Abstreiter},
  \citenamefont {Oulton}, \citenamefont {Tartakovskii}, \citenamefont
  {Mowbray}, \citenamefont {Skolnick}, \citenamefont {Liew}, \citenamefont
  {Cullis},\ and\ \citenamefont {Hopkinson}}]{fin04}%
  \BibitemOpen
  \bibfield  {author} {\bibinfo {author} {\bibfnamefont {J.~J.}\ \bibnamefont
  {Finley}}, \bibinfo {author} {\bibfnamefont {M.}~\bibnamefont {Sabathil}},
  \bibinfo {author} {\bibfnamefont {P.}~\bibnamefont {Vogl}}, \bibinfo {author}
  {\bibfnamefont {G.}~\bibnamefont {Abstreiter}}, \bibinfo {author}
  {\bibfnamefont {R.}~\bibnamefont {Oulton}}, \bibinfo {author} {\bibfnamefont
  {A.~I.}\ \bibnamefont {Tartakovskii}}, \bibinfo {author} {\bibfnamefont
  {D.~J.}\ \bibnamefont {Mowbray}}, \bibinfo {author} {\bibfnamefont {M.~S.}\
  \bibnamefont {Skolnick}}, \bibinfo {author} {\bibfnamefont {S.~L.}\
  \bibnamefont {Liew}}, \bibinfo {author} {\bibfnamefont {A.~G.}\ \bibnamefont
  {Cullis}}, \ and\ \bibinfo {author} {\bibfnamefont {M.}~\bibnamefont
  {Hopkinson}},\ }\Doi {10.1103/PhysRevB.70.201308} {\bibfield  {journal}
  {\bibinfo  {journal} {Phys. Rev. B},\ }\textbf {\bibinfo {volume} {70}},\
  \bibinfo {pages} {201308} (\bibinfo {year} {2004})}\BibitemShut {NoStop}%
\bibitem [{\citenamefont {Kowalik}\ \emph
  {et~al.}(2006){\natexlab{a}}\citenamefont {Kowalik}, \citenamefont {Krebs},
  \citenamefont {Senellart}, \citenamefont {Lema\^{i}tre}, \citenamefont
  {Eble}, \citenamefont {Kudelski}, \citenamefont {Gaj},\ and\ \citenamefont
  {Voisin}}]{kow06}%
  \BibitemOpen
  \bibfield  {author} {\bibinfo {author} {\bibfnamefont {K.}~\bibnamefont
  {Kowalik}}, \bibinfo {author} {\bibfnamefont {O.}~\bibnamefont {Krebs}},
  \bibinfo {author} {\bibfnamefont {P.}~\bibnamefont {Senellart}}, \bibinfo
  {author} {\bibfnamefont {A.}~\bibnamefont {Lema\^{i}tre}}, \bibinfo {author}
  {\bibfnamefont {B.}~\bibnamefont {Eble}}, \bibinfo {author} {\bibfnamefont
  {A.}~\bibnamefont {Kudelski}}, \bibinfo {author} {\bibfnamefont {J.~A.}\
  \bibnamefont {Gaj}}, \ and\ \bibinfo {author} {\bibfnamefont
  {P.}~\bibnamefont {Voisin}},\ }\href@noop {} {\bibfield  {journal} {\bibinfo
  {journal} {phys. stat. sol. c},\ }\textbf {\bibinfo {volume} {3}},\ \bibinfo
  {pages} {3890} (\bibinfo {year} {2006}{\natexlab{a}})}\BibitemShut {NoStop}%
\bibitem [{\citenamefont {Zrenner}\ \emph {et~al.}(2002)\citenamefont
  {Zrenner}, \citenamefont {Beham}, \citenamefont {Stufler}, \citenamefont
  {Findeis}, \citenamefont {Bichler},\ and\ \citenamefont
  {Abstreiter}}]{zre02}%
  \BibitemOpen
  \bibfield  {author} {\bibinfo {author} {\bibfnamefont {A.}~\bibnamefont
  {Zrenner}}, \bibinfo {author} {\bibfnamefont {E.}~\bibnamefont {Beham}},
  \bibinfo {author} {\bibfnamefont {S.}~\bibnamefont {Stufler}}, \bibinfo
  {author} {\bibfnamefont {F.}~\bibnamefont {Findeis}}, \bibinfo {author}
  {\bibfnamefont {M.}~\bibnamefont {Bichler}}, \ and\ \bibinfo {author}
  {\bibfnamefont {G.}~\bibnamefont {Abstreiter}},\ }\href@noop {} {\bibfield
  {journal} {\bibinfo  {journal} {Nature},\ }\textbf {\bibinfo {volume}
  {418}},\ \bibinfo {pages} {612} (\bibinfo {year} {2002})}\BibitemShut
  {NoStop}%
\bibitem [{\citenamefont {Seidl}\ \emph {et~al.}(2005)\citenamefont {Seidl},
  \citenamefont {Kroner}, \citenamefont {Dalgarno}, \citenamefont {Hogele},
  \citenamefont {Smith}, \citenamefont {Ediger}, \citenamefont {Gerardot},
  \citenamefont {Garcia}, \citenamefont {Petroff}, \citenamefont {Karrai},\
  and\ \citenamefont {Warburton}}]{sei05}%
  \BibitemOpen
  \bibfield  {author} {\bibinfo {author} {\bibfnamefont {S.}~\bibnamefont
  {Seidl}}, \bibinfo {author} {\bibfnamefont {M.}~\bibnamefont {Kroner}},
  \bibinfo {author} {\bibfnamefont {P.~A.}\ \bibnamefont {Dalgarno}}, \bibinfo
  {author} {\bibfnamefont {A.}~\bibnamefont {Hogele}}, \bibinfo {author}
  {\bibfnamefont {J.~M.}\ \bibnamefont {Smith}}, \bibinfo {author}
  {\bibfnamefont {M.}~\bibnamefont {Ediger}}, \bibinfo {author} {\bibfnamefont
  {B.~D.}\ \bibnamefont {Gerardot}}, \bibinfo {author} {\bibfnamefont {J.~M.}\
  \bibnamefont {Garcia}}, \bibinfo {author} {\bibfnamefont {P.~M.}\
  \bibnamefont {Petroff}}, \bibinfo {author} {\bibfnamefont {K.}~\bibnamefont
  {Karrai}}, \ and\ \bibinfo {author} {\bibfnamefont {R.~J.}\ \bibnamefont
  {Warburton}},\ }\Doi {10.1103/PhysRevB.72.195339} {\bibfield  {journal}
  {\bibinfo  {journal} {Phys. Rev. B},\ }\textbf {\bibinfo {volume} {72}},\
  \bibinfo {eid} {195339} (\bibinfo {year} {2005})}\BibitemShut {NoStop}%
\bibitem [{\citenamefont {Fry}\ \emph {et~al.}(2000)\citenamefont {Fry},
  \citenamefont {Itskevich}, \citenamefont {Mowbray}, \citenamefont {Skolnick},
  \citenamefont {Finley}, \citenamefont {Barker}, \citenamefont {O'Reilly},
  \citenamefont {Wilson}, \citenamefont {Larkin}, \citenamefont {Maksym},
  \citenamefont {Hopkinson}, \citenamefont {Al-Khafaji}, \citenamefont {David},
  \citenamefont {Cullis}, \citenamefont {Hill},\ and\ \citenamefont
  {Clark}}]{fry00}%
  \BibitemOpen
  \bibfield  {author} {\bibinfo {author} {\bibfnamefont {P.~W.}\ \bibnamefont
  {Fry}}, \bibinfo {author} {\bibfnamefont {I.~E.}\ \bibnamefont {Itskevich}},
  \bibinfo {author} {\bibfnamefont {D.~J.}\ \bibnamefont {Mowbray}}, \bibinfo
  {author} {\bibfnamefont {M.~S.}\ \bibnamefont {Skolnick}}, \bibinfo {author}
  {\bibfnamefont {J.~J.}\ \bibnamefont {Finley}}, \bibinfo {author}
  {\bibfnamefont {J.~A.}\ \bibnamefont {Barker}}, \bibinfo {author}
  {\bibfnamefont {E.~P.}\ \bibnamefont {O'Reilly}}, \bibinfo {author}
  {\bibfnamefont {L.~R.}\ \bibnamefont {Wilson}}, \bibinfo {author}
  {\bibfnamefont {I.~A.}\ \bibnamefont {Larkin}}, \bibinfo {author}
  {\bibfnamefont {P.~A.}\ \bibnamefont {Maksym}}, \bibinfo {author}
  {\bibfnamefont {M.}~\bibnamefont {Hopkinson}}, \bibinfo {author}
  {\bibfnamefont {M.}~\bibnamefont {Al-Khafaji}}, \bibinfo {author}
  {\bibfnamefont {J.~P.~R.}\ \bibnamefont {David}}, \bibinfo {author}
  {\bibfnamefont {A.~G.}\ \bibnamefont {Cullis}}, \bibinfo {author}
  {\bibfnamefont {G.}~\bibnamefont {Hill}}, \ and\ \bibinfo {author}
  {\bibfnamefont {J.~C.}\ \bibnamefont {Clark}},\ }\Doi
  {10.1103/PhysRevLett.84.733} {\bibfield  {journal} {\bibinfo  {journal}
  {Phys. Rev. Lett.},\ }\textbf {\bibinfo {volume} {84}},\ \bibinfo {pages}
  {733} (\bibinfo {year} {2000})}\BibitemShut {NoStop}%
\bibitem [{\citenamefont {Barker}\ and\ \citenamefont
  {O\char39{}Reilly}(2000)}]{bar00}%
  \BibitemOpen
  \bibfield  {author} {\bibinfo {author} {\bibfnamefont {J.~A.}\ \bibnamefont
  {Barker}}\ and\ \bibinfo {author} {\bibfnamefont {E.~P.}\ \bibnamefont
  {O\char39{}Reilly}},\ }\Doi {10.1103/PhysRevB.61.13840} {\bibfield  {journal}
  {\bibinfo  {journal} {Phys. Rev. B},\ }\textbf {\bibinfo {volume} {61}},\
  \bibinfo {pages} {13840} (\bibinfo {year} {2000})}\BibitemShut {NoStop}%
\bibitem [{\citenamefont {Feucker}\ \emph {et~al.}(2008)\citenamefont
  {Feucker}, \citenamefont {Seguin}, \citenamefont {Rodt}, \citenamefont
  {Hoffman},\ and\ \citenamefont {Bimberg}}]{feu08}%
  \BibitemOpen
  \bibfield  {author} {\bibinfo {author} {\bibfnamefont {M.}~\bibnamefont
  {Feucker}}, \bibinfo {author} {\bibfnamefont {R.}~\bibnamefont {Seguin}},
  \bibinfo {author} {\bibfnamefont {S.}~\bibnamefont {Rodt}}, \bibinfo {author}
  {\bibfnamefont {A.}~\bibnamefont {Hoffman}}, \ and\ \bibinfo {author}
  {\bibfnamefont {D.}~\bibnamefont {Bimberg}},\ }\href@noop {} {\bibfield
  {journal} {\bibinfo  {journal} {Appl. Phys. Lett.},\ }\textbf {\bibinfo
  {volume} {92}},\ \bibinfo {pages} {63116} (\bibinfo {year}
  {2008})}\BibitemShut {NoStop}%
\bibitem [{\citenamefont {Bellessa}\ \emph {et~al.}(1998)\citenamefont
  {Bellessa}, \citenamefont {Voliotis}, \citenamefont {Grousson}, \citenamefont
  {Wang}, \citenamefont {Ogura},\ and\ \citenamefont {Matsuhata}}]{bel98}%
  \BibitemOpen
  \bibfield  {author} {\bibinfo {author} {\bibfnamefont {J.}~\bibnamefont
  {Bellessa}}, \bibinfo {author} {\bibfnamefont {V.}~\bibnamefont {Voliotis}},
  \bibinfo {author} {\bibfnamefont {R.}~\bibnamefont {Grousson}}, \bibinfo
  {author} {\bibfnamefont {X.~L.}\ \bibnamefont {Wang}}, \bibinfo {author}
  {\bibfnamefont {M.}~\bibnamefont {Ogura}}, \ and\ \bibinfo {author}
  {\bibfnamefont {H.}~\bibnamefont {Matsuhata}},\ }\href@noop {} {\bibfield
  {journal} {\bibinfo  {journal} {Phys. Rev. B},\ }\textbf {\bibinfo {volume}
  {58}},\ \bibinfo {pages} {9933} (\bibinfo {year} {1998})}\BibitemShut
  {NoStop}%
\bibitem [{\citenamefont {Hours}\ \emph {et~al.}(2005)\citenamefont {Hours},
  \citenamefont {Senellart}, \citenamefont {Peter}, \citenamefont {Cavanna},\
  and\ \citenamefont {Bloch}}]{hou05}%
  \BibitemOpen
  \bibfield  {author} {\bibinfo {author} {\bibfnamefont {J.}~\bibnamefont
  {Hours}}, \bibinfo {author} {\bibfnamefont {P.}~\bibnamefont {Senellart}},
  \bibinfo {author} {\bibfnamefont {E.}~\bibnamefont {Peter}}, \bibinfo
  {author} {\bibfnamefont {A.}~\bibnamefont {Cavanna}}, \ and\ \bibinfo
  {author} {\bibfnamefont {J.}~\bibnamefont {Bloch}},\ }\href@noop {}
  {\bibfield  {journal} {\bibinfo  {journal} {Phys. Rev. B},\ }\textbf
  {\bibinfo {volume} {71}},\ \bibinfo {pages} {161306} (\bibinfo {year}
  {2005})}\BibitemShut {NoStop}%
\bibitem [{\citenamefont {Seufert}\ \emph {et~al.}(2003)\citenamefont
  {Seufert}, \citenamefont {Rambach}, \citenamefont {Bacher}, \citenamefont
  {Forchel}, \citenamefont {Passow},\ and\ \citenamefont {Hommel}}]{seu03}%
  \BibitemOpen
  \bibfield  {author} {\bibinfo {author} {\bibfnamefont {J.}~\bibnamefont
  {Seufert}}, \bibinfo {author} {\bibfnamefont {M.}~\bibnamefont {Rambach}},
  \bibinfo {author} {\bibfnamefont {G.}~\bibnamefont {Bacher}}, \bibinfo
  {author} {\bibfnamefont {A.}~\bibnamefont {Forchel}}, \bibinfo {author}
  {\bibfnamefont {T.}~\bibnamefont {Passow}}, \ and\ \bibinfo {author}
  {\bibfnamefont {D.}~\bibnamefont {Hommel}},\ }\Doi {10.1063/1.1580632}
  {\bibfield  {journal} {\bibinfo  {journal} {Appl. Phys. Lett.},\ }\textbf
  {\bibinfo {volume} {82}},\ \bibinfo {pages} {3946} (\bibinfo {year}
  {2003})}\BibitemShut {NoStop}%
\bibitem [{\citenamefont {Tinjod}\ \emph {et~al.}(2003)\citenamefont {Tinjod},
  \citenamefont {Gilles}, \citenamefont {Moehl}, \citenamefont {Kheng},\ and\
  \citenamefont {Mariette}}]{tin03}%
  \BibitemOpen
  \bibfield  {author} {\bibinfo {author} {\bibfnamefont {F.}~\bibnamefont
  {Tinjod}}, \bibinfo {author} {\bibfnamefont {B.}~\bibnamefont {Gilles}},
  \bibinfo {author} {\bibfnamefont {S.}~\bibnamefont {Moehl}}, \bibinfo
  {author} {\bibfnamefont {K.}~\bibnamefont {Kheng}}, \ and\ \bibinfo {author}
  {\bibfnamefont {H.}~\bibnamefont {Mariette}},\ }\Doi {10.1063/1.1583141}
  {\bibfield  {journal} {\bibinfo  {journal} {App. Phys. Lett.},\ }\textbf
  {\bibinfo {volume} {82}},\ \bibinfo {pages} {4340} (\bibinfo {year}
  {2003})}\BibitemShut {NoStop}%
\bibitem [{\citenamefont {Oulton}\ \emph {et~al.}(2002)\citenamefont {Oulton},
  \citenamefont {Finley}, \citenamefont {Ashmore}, \citenamefont {Gregory},
  \citenamefont {Mowbray}, \citenamefont {Skolnick}, \citenamefont {Steer},
  \citenamefont {Liew}, \citenamefont {Migliorato},\ and\ \citenamefont
  {Cullis}}]{oul02}%
  \BibitemOpen
  \bibfield  {author} {\bibinfo {author} {\bibfnamefont {R.}~\bibnamefont
  {Oulton}}, \bibinfo {author} {\bibfnamefont {J.~J.}\ \bibnamefont {Finley}},
  \bibinfo {author} {\bibfnamefont {A.~D.}\ \bibnamefont {Ashmore}}, \bibinfo
  {author} {\bibfnamefont {I.~S.}\ \bibnamefont {Gregory}}, \bibinfo {author}
  {\bibfnamefont {D.~J.}\ \bibnamefont {Mowbray}}, \bibinfo {author}
  {\bibfnamefont {M.~S.}\ \bibnamefont {Skolnick}}, \bibinfo {author}
  {\bibfnamefont {M.~J.}\ \bibnamefont {Steer}}, \bibinfo {author}
  {\bibfnamefont {S.-L.}\ \bibnamefont {Liew}}, \bibinfo {author}
  {\bibfnamefont {M.~A.}\ \bibnamefont {Migliorato}}, \ and\ \bibinfo {author}
  {\bibfnamefont {A.~J.}\ \bibnamefont {Cullis}},\ }\Doi
  {10.1103/PhysRevB.66.045313} {\bibfield  {journal} {\bibinfo  {journal}
  {Phys. Rev. B},\ }\textbf {\bibinfo {volume} {66}},\ \bibinfo {pages}
  {045313} (\bibinfo {year} {2002})}\BibitemShut {NoStop}%
\bibitem [{\citenamefont {Franceschetti}\ and\ \citenamefont
  {Zunger}(2000)}]{fra00}%
  \BibitemOpen
  \bibfield  {author} {\bibinfo {author} {\bibfnamefont {A.}~\bibnamefont
  {Franceschetti}}\ and\ \bibinfo {author} {\bibfnamefont {A.}~\bibnamefont
  {Zunger}},\ }\Doi {10.1103/PhysRevB.62.2614} {\bibfield  {journal} {\bibinfo
  {journal} {Phys. Rev. B},\ }\textbf {\bibinfo {volume} {62}},\ \bibinfo
  {pages} {2614} (\bibinfo {year} {2000})}\BibitemShut {NoStop}%
\bibitem [{\citenamefont {Nguyen}\ \emph {et~al.}(2007)\citenamefont {Nguyen},
  \citenamefont {Mackowski}, \citenamefont {Hoang}, \citenamefont {Jackson},
  \citenamefont {Smith},\ and\ \citenamefont {Karczewski}}]{ngu07}%
  \BibitemOpen
  \bibfield  {author} {\bibinfo {author} {\bibfnamefont {T.~A.}\ \bibnamefont
  {Nguyen}}, \bibinfo {author} {\bibfnamefont {S.}~\bibnamefont {Mackowski}},
  \bibinfo {author} {\bibfnamefont {T.~B.}\ \bibnamefont {Hoang}}, \bibinfo
  {author} {\bibfnamefont {H.~E.}\ \bibnamefont {Jackson}}, \bibinfo {author}
  {\bibfnamefont {L.~M.}\ \bibnamefont {Smith}}, \ and\ \bibinfo {author}
  {\bibfnamefont {G.}~\bibnamefont {Karczewski}},\ }\Doi
  {10.1103/PhysRevB.76.245320} {\bibfield  {journal} {\bibinfo  {journal}
  {Phys. Revi. B},\ }\textbf {\bibinfo {volume} {76}},\ \bibinfo {eid} {245320}
  (\bibinfo {year} {2007})}\BibitemShut {NoStop}%
\bibitem [{\citenamefont {Kazimierczuk}\ \emph {et~al.}(2009)\citenamefont
  {Kazimierczuk}, \citenamefont {Suffczynski}, \citenamefont {Golnik},
  \citenamefont {Gaj}, \citenamefont {Kossacki},\ and\ \citenamefont
  {Wojnar}}]{kaz09}%
  \BibitemOpen
  \bibfield  {author} {\bibinfo {author} {\bibfnamefont {T.}~\bibnamefont
  {Kazimierczuk}}, \bibinfo {author} {\bibfnamefont {J.}~\bibnamefont
  {Suffczynski}}, \bibinfo {author} {\bibfnamefont {A.}~\bibnamefont {Golnik}},
  \bibinfo {author} {\bibfnamefont {J.~A.}\ \bibnamefont {Gaj}}, \bibinfo
  {author} {\bibfnamefont {P.}~\bibnamefont {Kossacki}}, \ and\ \bibinfo
  {author} {\bibfnamefont {P.}~\bibnamefont {Wojnar}},\ }\Doi
  {10.1103/PhysRevB.79.153301} {\bibfield  {journal} {\bibinfo  {journal}
  {Phys. Rev. B},\ }\textbf {\bibinfo {volume} {79}},\ \bibinfo {eid} {153301}
  (\bibinfo {year} {2009})}\BibitemShut {NoStop}%
\bibitem [{\citenamefont {Loudon}(1973)}]{lou73}%
  \BibitemOpen
  \bibfield  {author} {\bibinfo {author} {\bibfnamefont {R.}~\bibnamefont
  {Loudon}},\ }\href@noop {} {\emph {\bibinfo {title} {The Quantum Theory of
  Light}}}\ (\bibinfo  {publisher} {Oxford},\ \bibinfo {year}
  {1973})\BibitemShut {NoStop}%
\bibitem [{\citenamefont {Knox}(1963)}]{kno63}%
  \BibitemOpen
  \bibfield  {author} {\bibinfo {author} {\bibfnamefont {R.~S.}\ \bibnamefont
  {Knox}},\ }\href@noop {} {\emph {\bibinfo {title} {Theory of Excitons}}}\
  (\bibinfo  {publisher} {Academic Press, New York},\ \bibinfo {year}
  {1963})\BibitemShut {NoStop}%
\bibitem [{\citenamefont {Bryant}(1988)}]{bry88}%
  \BibitemOpen
  \bibfield  {author} {\bibinfo {author} {\bibfnamefont {G.~W.}\ \bibnamefont
  {Bryant}},\ }\href@noop {} {\bibfield  {journal} {\bibinfo  {journal} {Phys.
  Rev. B},\ }\textbf {\bibinfo {volume} {37}},\ \bibinfo {pages} {8763}
  (\bibinfo {year} {1988})}\BibitemShut {NoStop}%
\bibitem [{\citenamefont {Corni}\ \emph {et~al.}(2003)\citenamefont {Corni},
  \citenamefont {Br\'{a}sken}, \citenamefont {Lindberg}, \citenamefont {Olsen},\ and\
  \citenamefont {Sundholm}}]{cor03}%
  \BibitemOpen
  \bibfield  {author} {\bibinfo {author} {\bibfnamefont {S.}~\bibnamefont
  {Corni}}, \bibinfo {author} {\bibfnamefont {M.}~\bibnamefont {Br\'{a}sken}},
  \bibinfo {author} {\bibfnamefont {M.}~\bibnamefont {Lindberg}}, \bibinfo {author}
  {\bibfnamefont {J.}~\bibnamefont {Olsen}}, \ and\ \bibinfo {author}
  {\bibfnamefont {D.}~\bibnamefont {Sundholm}},\ }\href@noop {} {\bibfield
  {journal} {\bibinfo  {journal} {Phys. Rev. B},\ }\textbf {\bibinfo {volume}
  {67}},\ \bibinfo {pages} {045313} (\bibinfo {year} {2003})}\BibitemShut
  {NoStop}%
\bibitem [{\citenamefont {Suffczy\'{n}ski}\ \emph {et~al.}(2006)\citenamefont
  {Suffczy\'{n}ski}, \citenamefont {Kazimierczuk}, \citenamefont {Goryca},
  \citenamefont {Piechal}, \citenamefont {Trajnerowicz}, \citenamefont
  {Kowalik}, \citenamefont {Kossacki}, \citenamefont {Golnik}, \citenamefont
  {Korona}, \citenamefont {Nawrocki}, \citenamefont {Gaj},\ and\ \citenamefont
  {Karczewski}}]{suf06}%
  \BibitemOpen
  \bibfield  {author} {\bibinfo {author} {\bibfnamefont {J.}~\bibnamefont
  {Suffczy\'{n}ski}}, \bibinfo {author} {\bibfnamefont {T.}~\bibnamefont
  {Kazimierczuk}}, \bibinfo {author} {\bibfnamefont {M.}~\bibnamefont
  {Goryca}}, \bibinfo {author} {\bibfnamefont {B.}~\bibnamefont {Piechal}},
  \bibinfo {author} {\bibfnamefont {A.}~\bibnamefont {Trajnerowicz}}, \bibinfo
  {author} {\bibfnamefont {K.}~\bibnamefont {Kowalik}}, \bibinfo {author}
  {\bibfnamefont {P.}~\bibnamefont {Kossacki}}, \bibinfo {author}
  {\bibfnamefont {A.}~\bibnamefont {Golnik}}, \bibinfo {author} {\bibfnamefont
  {K.~P.}\ \bibnamefont {Korona}}, \bibinfo {author} {\bibfnamefont
  {M.}~\bibnamefont {Nawrocki}}, \bibinfo {author} {\bibfnamefont {J.~A.}\
  \bibnamefont {Gaj}}, \ and\ \bibinfo {author} {\bibfnamefont
  {G.}~\bibnamefont {Karczewski}},\ }\Doi {10.1103/PhysRevB.74.085319}
  {\bibfield  {journal} {\bibinfo  {journal} {Phys. Rev. B},\ }\textbf
  {\bibinfo {volume} {74}},\ \bibinfo {eid} {085319} (\bibinfo {year}
  {2006})}\BibitemShut {NoStop}%
\bibitem [{\citenamefont {Kowalik}\ \emph
  {et~al.}(2006){\natexlab{b}}\citenamefont {Kowalik}, \citenamefont
  {Kudelski}, \citenamefont {Golnik}, \citenamefont {Suffczyński},
  \citenamefont {Krebs}, \citenamefont {Voisin}, \citenamefont {Karczewski},
  \citenamefont {Kossut},\ and\ \citenamefont {Gaj}}]{kow06b}%
  \BibitemOpen
  \bibfield  {author} {\bibinfo {author} {\bibfnamefont {K.}~\bibnamefont
  {Kowalik}}, \bibinfo {author} {\bibfnamefont {A.}~\bibnamefont {Kudelski}},
  \bibinfo {author} {\bibfnamefont {A.}~\bibnamefont {Golnik}}, \bibinfo
  {author} {\bibfnamefont {J.}~\bibnamefont {Suffczyński}}, \bibinfo {author}
  {\bibfnamefont {O.}~\bibnamefont {Krebs}}, \bibinfo {author} {\bibfnamefont
  {P.}~\bibnamefont {Voisin}}, \bibinfo {author} {\bibfnamefont
  {G.}~\bibnamefont {Karczewski}}, \bibinfo {author} {\bibfnamefont
  {J.}~\bibnamefont {Kossut}}, \ and\ \bibinfo {author} {\bibfnamefont {J.~A.}\
  \bibnamefont {Gaj}},\ }\href@noop {} {\bibfield  {journal} {\bibinfo
  {journal} {phys. stat. sol. c},\ }\textbf {\bibinfo {volume} {3}},\ \bibinfo
  {pages} {865} (\bibinfo {year} {2006}{\natexlab{b}})}\BibitemShut {NoStop}%
\bibitem [{\citenamefont {Riva}\ \emph {et~al.}(2000)\citenamefont {Riva},
  \citenamefont {Peeters},\ and\ \citenamefont {Varga}}]{riv00}%
  \BibitemOpen
  \bibfield  {author} {\bibinfo {author} {\bibfnamefont {C.}~\bibnamefont
  {Riva}}, \bibinfo {author} {\bibfnamefont {F.~M.}\ \bibnamefont {Peeters}}, \
  and\ \bibinfo {author} {\bibfnamefont {K.}~\bibnamefont {Varga}},\ }\Doi
  {10.1103/PhysRevB.61.13873} {\bibfield  {journal} {\bibinfo  {journal} {Phys.
  Rev. B},\ }\textbf {\bibinfo {volume} {61}},\ \bibinfo {pages} {13873}
  (\bibinfo {year} {2000})}\BibitemShut {NoStop}%
\end{thebibliography}
\end{document}